\documentclass[12pt]{article}

\usepackage[linktocpage=true,plainpages=false]{hyperref}
\usepackage{color}
\usepackage{amsmath,amssymb}
\usepackage{cite}
\usepackage{graphicx}
\usepackage{array}
\usepackage{multicol}
\usepackage[T1]{fontenc}
\usepackage[labelsep=period]{caption}

\definecolor{linkcolor}{rgb}{0.6,0,0}
\definecolor{citecolor}{rgb}{0,0.6,0}
\definecolor{urlcolor}{rgb}{0,0,0.9}
\hypersetup{colorlinks, linkcolor={linkcolor},citecolor={citecolor}, urlcolor={urlcolor}}

\newcommand{\zagt}[1]{\hbox to 4em{\hfil#1\hfil}}
\newcommand{\str}[1]{\mathrel{\mathop{\longrightarrow}\limits_{#1}}}
\newcommand{\dd}{\partial}

\newcommand{\m}{\mu}
\newcommand{\n}{\nu}
\newcommand{\ls}{\left(}
\newcommand{\rs}{\right)}
\newcommand{\al}{\alpha}

\newcommand{\ff}{\varphi}

\newcommand{\La}{\Lambda}

\newcommand{\te}{\theta}

\newcommand{\dz}{\zeta}
\newcommand{\si}{\sigma}
\newcommand{\sign}{{\rm sign}}

\newcommand{\disn}[2]{$$\displaylines{\refstepcounter{equation}%
            \label{#1}\hskip 1em minus 1em #2\hfilneg}$$}
\newcommand{\nom}{\hfil\hskip 1em minus 1em (\theequation)}
\newcommand{\no}{\hfil \hskip 1em minus 1em\phantom{(\theequation)}%
            \hfilneg\cr\hfilneg\hskip 1em minus 1em\hfil}

\begin{document}
\title{Classification of global minimal embeddings\\
for nonrotating black holes}
\author{A.~A.~Sheykin\thanks{E-mail: a.sheykin@spbu.ru}, S.~A.~Paston\thanks{E-mail: paston@pobox.spbu.ru}
\\
{\it Saint Petersburg State University, Saint Petersburg, Russia}
}
\date{\vskip 15mm}
\maketitle

\begin{abstract}
We consider the problem of the existence of global embeddings of metrics of spherically symmetric black
holes into an ambient space with the minimal possible dimension. We classify the possible types of
embeddings by the type of realization of the metric symmetry by ambient space symmetries. For the
Schwarzschild, Schwarzschild-de Sitter, and Reissner-Nordstr$\ddot{\text{o}}$m black holes, we prove that the known
global embeddings are the only ones. We obtain a new global embedding for the Reissner-Nordstr$\ddot{\text{o}}$m-de Sitter metrics and prove that constructing such embeddings is impossible for the Schwarzschild-anti-de Sitter metric. We also discuss the possibility of constructing global embeddings of the Reissner-Nordstr$\ddot{\text{o}}$m-anti-de Sitter metric.
\end{abstract}

\newpage

\section{Introduction}
According to the Friedman theorem \cite{fridman61}, an arbitrary $n$-dimensional pseudo-Riemannian
manifold can always be locally isometrically embedded in a flat space of $N\geqslant n(n+1)/2$ dimensions. In
some cases, most often in the presence of some symmetry, the sufficient value of N can turn out to be
smaller \cite{schmutzer}, and the embedding can be not only local but also global. By an embedding, we mean specifying
a surface in a flat ambient space that is described by the embedding functions
 $y^a(x^\m)$, whose induced metric
\disn{1}{
g_{\m\n}=(\dd_\m y^a) (\dd_\n y^b) \eta_{ab}
\nom}
reproduces a specified metric of the embedded pseudo-Riemannian manifold (here and hereafter, the sub-
scripts $\m,\n,\ldots$  label its coordinates, the superscripts $a,b,\ldots$ label Lorentz coordinates in the ambient
space, and $\eta_{ab}$ is the flat metric of the ambient space).

Because our space-time is a pseudo-Riemannian manifold in the framework of general relativity, we
can perform the embedding procedure for it and describe it as a four-dimensional surface in a flat ambient
space. It suffices to take ten as its dimension N, and it usually turns out to be smaller for a specific metric
with symmetries (but we note that larger values are used in some approaches, for example,  $N=14$ \cite{willison}).

Embeddings with the smallest ambient-space dimension $N_{min}$ for a given metric are said to be minimal.
Embeddings began to be used in general relativity almost immediately after it appeared. In particular,
the first embedding of the Schwarzschild metric was constructed in 1921\cite{kasner3}, only five years after the discovery
of the metric itself. The embedding can be used for various purposes in general relativity. Embeddings
were mainly used to classify exact solutions up to the 1970s \cite{schmutzer} because the embedding class $p=N_{min}-n$
is invariant and were also used to study the geometric properties of metrics. For example, the well-known Kruskal coordinates for the Schwarzschild metric were most likely obtained using the Fronsdal embedding\cite{frons}.

The idea of using the embedding function $y^a(x)$  as a dynamical variable instead of the metric $g_{\m\n}(x)$ to
describe gravity was first proposed in \cite{regge}. Different variants of such a description of gravity as an embedding
theory were studied more than once afterwards (see, e.g., \cite{deser,pavsic85let,tapia,maia89,bandos,davkar,statja18,faddeev,statja26}), sometimes appearing independently.

Explicit embeddings of physically interesting solutions of the Einstein equations can be useful for more
than just analyzing the structure of a resulting manifold. There exist efforts to establish a relation between
quantum effects produced in the Riemannian space and in a flat ambient space on the embedded surface.
For example, the correspondence between the Hawking and Unruh effects was discussed when considering
embeddings \cite{deserlev99}  (also see \cite{statja34,statja36} and the references therein). As a recent example of such an application of explicit embeddings, we mention \cite{1305.6933}, where the Green's function in the Riemannian space was calculated
using embeddings.

We note that metrics with $p\leq2$ can be studied relatively easily. Because all spherically symmetric
black holes have the embedding class $2$ (following \cite{karmarkar}, we can verify that  $p\neq1$ for them and $p=2$ for
the spherically symmetric metric of general form), and rather many embeddings are currently known for
them. In addition to \cite{kasner3,frons}, mentioned above, embeddings of the Schwarzschild metric in a six-dimensional
space were also proposed in \cite{fudjitani,davidson,statja27}, and it's worth noting that all possible minimal embeddings admitted by the metric symmetry were found in \cite{statja27}. Among six resulting embeddings, four turn out to be global, i.~e., smooth for
all radius values, and two others (proposed in \cite{kasner3,fudjitani}) cannot be continued under the horizon.

Minimal embeddings for the metric of the Reissner-Nordstr$\ddot{\text{o}}$m charged black hole were first proposed
in \cite{rosen65,RosenRN}, but they are not global in the most interesting case of the presence of two horizons (the so-called
nonextremal case). The complete set of symmetric global minimal embeddings for the nonextremal case
was found in \cite{statja30}.

In addition to the abovementioned black holes, among the spherically symmetric ones, uncharged
and charged black holes in the de Sitter and anti-de Sitter spaces are also known (we use the notation
S-dS and S-AdS for uncharged ones and RN-dS and RN-AdS for charged ones). For them, there is only
fragmentary information on the existence of global embeddings (local embeddings are easily obtained by
a simple generalization of the well-known embeddings of the Schwarzschild metric). For example, seven-
dimensional (consequently, not minimal) global embeddings for S-AdS and RN-AdS black holes were
proposed in \cite{gr-qc/0001045}. Minimal global embeddings for the S-dS black hole were found in \cite{statja32}. Embeddings
for axisymmetric black holes, i.~e., rotating Kerr and Kerr-Newman black holes, are hardly studied. There
exist only the Kuzeev embeddings with $N=9$ for Kerr \cite{kuzeev} and Kerr-Newman \cite{kuzeevRN} metrics, the 14-dimensional
embedding of the Kerr metric proposed in \cite{gr-qc/0503079}, and also different embedding diagrams
(embeddings of submanifolds, for example, with $\phi=0$), in which we are not interested now.

Here, we study the problem of the existence of global minimal (i.~e., six-dimensional) embeddings for
all nonrotating black holes, considering the nonextremal case corresponding to a maximum possible number
of horizons.

\section{Metrics under study and types of embeddings}\label{tipi}
An arbitrary spherically symmetric metric invariant under shifts of the parameter $t$ (i.~e., static in the
domains with the ordinary causality properties in which a shift of $t$ corresponds to a timelike interval)
can be written in the form
\disn{2}{
ds^2=g_{00}(r)dt^2+g_{11}(r)dr^2-r^2\ls d\te^2+\sin^2\!\te\, d\ff^2\rs.
\nom}
The metric components

\disn{3}{
g_{00}(r)=1-\frac{2m}{r}+\frac{q^2}{r^2}-\frac{\La r^2}{3}\equiv F(r),\qquad
g_{11}(r)=-\frac{1}{F(r)},
\nom}
correspond to the general case of a nonrotating black hole. Here, $m>0$ is the black hole mass, $q$ is its
charge, and $\Lambda$ is the cosmological constant, which is positive if the black hole is considered in the de Sitter
space and negative if it is considered in the anti-de Sitter space.

We study embeddings for six types of nonrotating black holes:
\begin{itemize}
\item Schwarzschild (S),
\item Schwarzschild-de Sitter (S-dS),
\item Schwarzschild-anti-de Sitter (S-AdS),
\item Reissner-Nordstr$\ddot{\text{o}}$m (RN),
\item Reissner-Nordstr$\ddot{\text{o}}$m-de Sitter (RN-dS), and
\item Reissner-Nordstr$\ddot{\text{o}}$m-anti-de Sitter (RN-AdS).
\end{itemize}
We restrict ourself to studying ranges of the parameters $m$, $q$ and $\Lambda$, such that the black hole is not
extremal, i.~e., has the maximum possible number of horizons. This corresponds to the case where the charge
$q$ and the cosmological constant $\La$ are rather small. The possible values of the parameters $q$, $\La$ and the
number of horizons  $n_{{\rm h}}$ for such black holes are given in Tab. \ref{tab1}.
\begin{table}[htbp]
\begin{center}
\begin{tabular}{|c|c|c|c|c|c|c|}
\hline
                &\zagt{S}&\zagt{S-dS}&\zagt{S-AdS}&\zagt{RN}&\zagt{RN-dS}&\zagt{RN-AdS}\\
\hline
$q$             & 0      &   0       &  0         & $\ne0$  & $\ne0$     &  $\ne 0$    \\
\hline
$\La$           & 0      &   >0      &    <0      &   0     &    >0      &   <0        \\
\hline
$n_{{\rm h}}$ & 1      &   2       &    1       &   2     &    3       &   2        \\
\hline
\end{tabular}
\end{center}
\caption{The values of the parameters $q$, $\La$ and $n_{{\rm h}}$ for the considered black holes.}
\label{tab1}
\end{table}

\begin{figure}[ht]
\begin{minipage}[b]{0.45\linewidth}
\centering
\includegraphics[width=\textwidth]{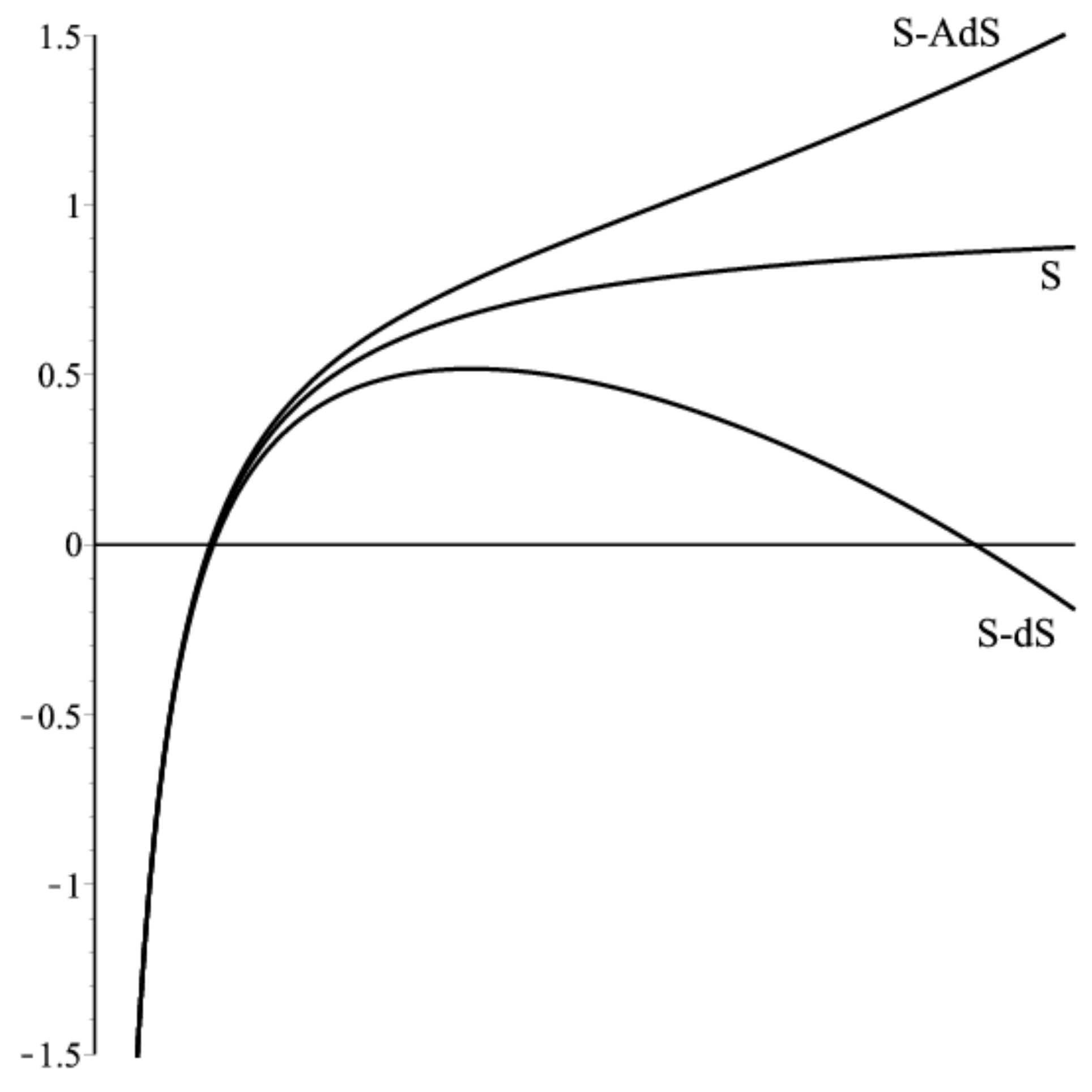}
\end{minipage}
\hspace{0.5cm}
\begin{minipage}[b]{0.45\linewidth}
\centering
\includegraphics[width=\textwidth]{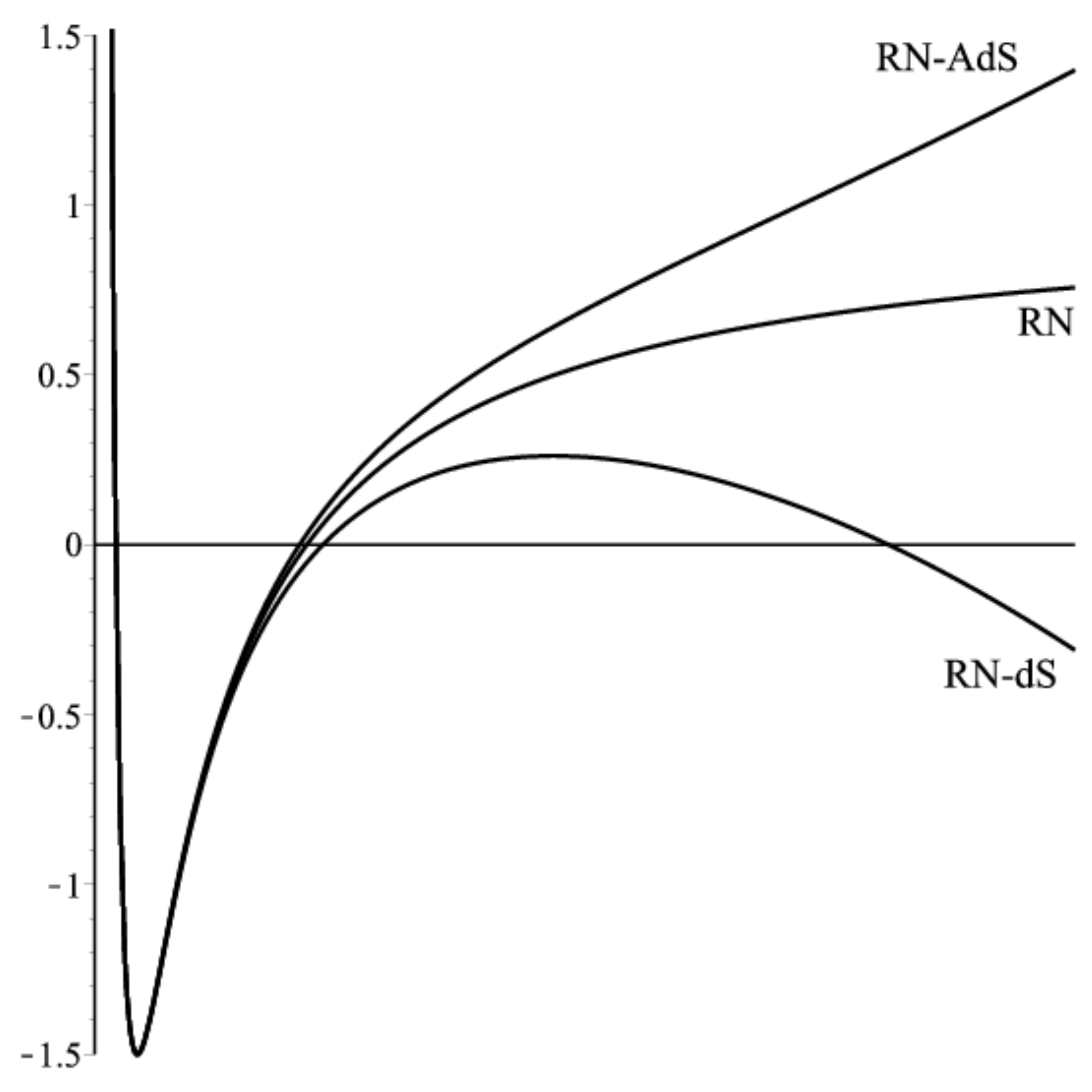}
\end{minipage}
\caption{\label{fig1}
Graphs of the functions $g_{00}(r)$ for the considered black holes.}
\end{figure}

The plots of the functions $g_{00}(r)$ for the considered black holes are shown in Fig.~\ref{fig1}.

All the six-dimensional embeddings (i.~e., four-dimensional surfaces in the flat six-dimen\-sional space)
having the symmetry of nonrotating black holes (the symmetry $SO(3)\times T^1$ ) were classified in \cite{statja27}.

We write the resulting nine types of embeddings, indicating allowed signatures of the ambient space for each
type.

Elliptic embeddings, the signature $\{\si,\si,\dz,-1,-1,-1\}$ (here and hereafter $\dz,\si,\rho=\pm1$):
 \disn{4}{
y^0=f(r)\sin(\al t+w(r)),\qquad
y^1=f(r)\cos(\al t+w(r)),\qquad
y^2=h(r),\no
y^3=r\,\cos\te,\qquad
y^4=r\,\sin\te\,\cos\ff,\qquad
y^5=r\,\sin\te\,\sin\ff.
\nom}

Parabolic embeddings, the signature $\{\dz,-\dz,-\dz,-1,-1,-1\}$:
 \disn{8}{
y^+=\frac{\alpha^2 t^2}{2}h(r)+\alpha f(r)t+w(r),\qquad
y^2=f(r)+ \alpha h(r) t,\qquad
y^-=h(r)
\nom}
(if the expressions for the components  $y^3,y^4,y^5$ are not given in this and subsequent embeddings, then
they coincide with those given in (\ref{4})).

Hyperbolic embeddings, the signature $\{1,-1,\dz,-1,-1,-1\}$:
 \disn{5}{
y^0=f(r)\sinh(\al t+w(r)),\qquad
y^1=f(r)\cosh(\al t+w(r)),\qquad
y^2=h(r)
\nom}
or, for other values of $r$:
 \disn{5.1}{
y^0=f(r)\cosh(\al t+w(r)),\qquad
y^1=f(r)\sinh(\al t+w(r)),\qquad
y^2=h(r).
\nom}
When considering embeddings of this type, we use the function $\xi(r)$, which we set to $+1$ if formula (\ref{5}),
is used for a given $r$ and $-1$ if formula \eqref{5.1} is used.

Spiral embeddings, the signature $\{\si,\si,\dz,-1,-1,-1\}$:
 \disn{6}{
y^0=f(r)\sin(\alpha t+w(r)),\qquad
y^1=f(r)\cos(\alpha t+w(r)),\qquad
y^2=k t+h(r).
\nom}

Exponential embeddings, the signature $\{1,-1,\dz,-1,-1,-1\}$:
 \disn{7}{
y^+=\frac{1}{\al}\exp(\alpha t+w(r)),\qquad
y^-=f(r)\exp(-(\alpha t+w(r))),\qquad
y^2=k t+h(r),
\nom}
where $y^\pm=(y^0\pm y^1)/\sqrt{2}$. The signature is indicated for the Lorentz components $y^0,y^1,\ldots$.

Cubic embeddings, the signature $\{\dz,-\dz,-\dz,-1,-1,-1\}$:
 \disn{9}{
y^+=\frac{\alpha^2 \widetilde t^3}{6}+\alpha f(r)\widetilde t+u(r),\qquad
y^2=\frac{\alpha \widetilde t^2}{2}+f(r),\qquad
y^-=\widetilde t,
\nom}
where $\widetilde t=t+h(r)$.

In addition to the abovementioned six embeddings, there are three types of embeddings with the required symmetry, but their freedom is too small to use them when seeking embeddings of specific metrics. To
make the picture more complete, we also write them: the embedding with the signature $\{\dz,\si,\rho,-1,-1,-1\}$:
 \disn{10}{
y^0=k_0 t+h_0(r),\quad
y^1=k_1 t +h_1(r),\quad
y^2=k_2 t+ h_2(r),
\nom}
the embedding with the signature $\{+1,+1,+1,-1,-1,-1\}$:
 \disn{12}{
\begin{array}{lcl}
y^0=r\sinh(\al t+w(r))\,\cos\te,                   &\quad &  y^3=r\cosh(\al t+w(r))\,\cos\te,  \\
y^1=r\sinh(\al t+w(r))\,\sin\te\,\cos\ff,          &\quad &  y^4=r\cosh(\al t+w(r))\,\sin\te\,\cos\ff,\\
y^2=r\sinh(\al t+w(r))\,\sin\te\,\sin\ff,          &\quad &  y^5=r\cosh(\al t+w(r))\,\sin\te\,\sin\ff\\
\end{array}
\nom}
and an embedding with the signature  $\{-1,-1,-1,-1,-1,-1\}$  that is analogous to (\ref{12})  in which sinh and
cosh are replaced with sin and cos (this embedding is not suitable for our purposes because of the Euclidean
signature). In the above formulas $\al,k$ --  are positive constants, and $f(r),w(r),h(r)$ are arbitrary
functions.
In what follows, we analyze the possibility of constructing global embeddings (\ref{4})-(\ref{12}) for the above
metrics.

\section{Analysis of the existence of global embeddings}\label{analiz}
We study the possibility of constructing global embeddings (i.~e., smooth for any values of $r$)  for the
above types of nonrotating black holes. Some of the above types of embeddings are not suitable for this
purpose, because the component $g_{00}$ of the metric induced by embeddings of these types cannot change
the sign, while all the metrics that are interesting to us have at least one horizon in which the sign of this
component changes (see Fig. 1).

The sign of the component $g_{00}$ turns out to be constant for embeddings of the elliptic type, for which
substituting (\ref{4}) in (\ref{1})  gives $g_{00}=\si\al^2 f(r)^2$,  of the parabolic type, where $g_{00}=-\dz\al^2 h(r)^2$; and also of type (\ref{10}), where $g_{00}=\dz k_0^2+\si k_1^2+\rho k_2^2=const$ and of type (\ref{12}), where $g_{00}=\al^2 r^2$. In what follows, we study the remaining four types of embeddings, namely, embeddings (\ref{5})-(\ref{9}).

\subsection{Hyperbolic embeddings}\label{hyp}
 We take into account that for hyperbolic embeddings given by formulas (\ref{5}),(\ref{5.1}),
 the choice of one of these two formulas is determined by the sign function $\xi(r)$.
Using (\ref{1}), we calculate the metric components
\disn{13}{
g_{00}=\xi\al^2 f^2,\qquad
g_{01}=\xi\al f^2w',\qquad
g_{11}=-\xi f'^2+\xi f^2 w'^2+\dz h'^2-1.
\nom}
Hence, using (\ref{2}) and (\ref{3}), we conclude that
\disn{14}{
\xi=\sign(F),\qquad
f=\frac{\sqrt{|F|}}{\al},\qquad
w'=0,
\nom}
and also obtain the equation
 \disn{15}{
\dz h'^2=\frac{1}{4\al^2F}\ls F'^2-4\al^2\rs+1.
\nom}
resulting from the formula for $g_{11}$.

 Because the function $F(r)$ vanishes at a certain point $r_*$ , $h(r)$ must
be smooth for the sought global embedding (because it coincides with the component $y^2$ of the embedding
function); it follows from Eq. (\ref{15}) that $F'(r_*)=\pm 2\al$. This means that the function $F(r)$ vanishes either
at only one point or at all points where this occurs; it has the same modulus of the derivative. As can be
seen from (\ref{3}), the latter does not occur in the general case of the black hole parameters. Therefore, the
possibility of constructing a global hyperbolic embedding can exist only for the S and S-AdS black holes (see
Table \ref{tab1}). For the Schwarzschild metric, such an embedding is known, namely, the Fronsdal embedding \cite{frons}.
We note that the passage of the function $F(r)$ through zero with the sign change does not lead to the
violation of the embedding function smoothness at the horizon, although $f(r)$ and $\xi(r)$ are not smooth
(see \cite{frons,statja27} for the details).

We study the S-AdS case. From Eq. (\ref{15}), if we take the asymptotic behavior of its right-hand side
into account, then we can find that
\disn{16}{
\dz h'^2 \str{r\to 0} -\frac{m}{2\al^2r^3} <0 ,\qquad \dz h'^2 \str{r\to \infty} \frac{|\La|}{3\al^2}+1>0.
\nom}
It follows from the second asymptotic form that $\dz=+1$, but this is inconsistent with the first asymptotic
form. Hence, a global hyperbolic embedding cannot be constructed for the S-AdS black hole.
As a result, we find that the minimal hyperbolic global embedding only exists for the Schwarzschild
black hole.

\subsection{Spiral embeddings}\label{spir}
Using (\ref{1}), we calculate the metric components
\disn{17}{
g_{00}=\si\al^2 f^2+\dz k^2,\qquad
g_{01}=\si\al f^2w'+\dz k h',\qquad
g_{11}=\si f'^2+\si f^2 w'^2+\dz h'^2-1,
\nom}
whence using (\ref{2}) and (\ref{3}) again, we conclude that
\disn{18}{
\si=\sign(F-\dz k^2),\qquad
f=\frac{\sqrt{|F-\dz k^2|}}{\al},\qquad
w'=-\frac{\dz\al k h'}{F-\dz k^2}.
\nom}
Because $\si$ and $\dz$ are constants, it hence follows that $F(r)$ must be bounded either above or below and the
quantity $F-\dz k^2$ therefore has a constant sign. The S-AdS and RN-dS black holes have no such property;
therefore, a global embedding cannot be constructed for them. The minimal global embeddings for the
S, S-dS, and RN black holes were respectively constructed in \cite{statja27}, \cite{statja32} and \cite{statja30}. It only remains for us to
analyze the possibility of constructing such an embedding in the RN-AdS case, which we now do.

The form of the function $F(r)$ in the RN-AdS case shows that to provide a definite sign of the quantity
$F-\dz k^2$, we must take $\dz=-1$ and $k^2\ge |F_{{\rm min}}|$ where $F_{{\rm min}}<0$ is the minimum of $F(r)$. We note that if
$k^2=|F_{{\rm min}}|$ is taken, then the resulting embedding contains a singularity at the point where $F(r)=F_{{\rm min}}$.
We therefore assume that $k^2>|F_{{\rm min}}|$. For the component $g_{11}$, we can obtain the equation
\disn{19}{
h'^2=\frac{F F'^2-4\al^2(F +k^2)(F-1)}{4\al^2F^2}.
\nom}
from formula (\ref{1}) ($\dz=-1$ is already set in it). Unlike the function $h(r)$ in Eq. (\ref{15}), the same function
in this equation tends to infinity at the horizons because the function $F(r)$ in the denominator vanishes.
This does not necessarily mean that the embedding function is not smooth at the horizons, because the
function $h(r)$ enters the component of embedding function  (\ref{6}) in the form of the sum with $kt$, and if the
external-observer time $t$ is used, there is a coordinate singularity of the metric: falling over the horizon
occurs only during an infinite time $t$.

The smoothness of the embedding function can be verified by passing to the delay time $\widetilde t=t+h(r)/k$
(in terms of which $y^2=k\widetilde t$). If it is used in (\ref{6}), then the smoothness of the component $y^0$ means that the
function $u(r)=w(r)-\al h(r)/k$ is smooth. If (\ref{18}) is used, then Eq. (\ref{19}) can be rewritten in terms of $u(r)$
as
\disn{20}{
u'^2=\frac{F F'^2-4\al^2(F +k^2)(F-1)}{4k^2(F +k^2)^2}.
\nom}
Because the denominator in this equation does not vanish anywhere, the embedding function is smooth if
the right-hand side of the equation is nonnegative for all $r$. To verify this, we first find its asymptotic form.
For the function $F(r)$ corresponding to the RN-AdS case, we find
\disn{20.1}{
u'^2\str{r\to 0}\frac{q^2}{k^2r^4}>0,\qquad
u'^2\str{r\to \infty}\frac{|\La|-3\al^2}{3k^2}
\nom}
A necessary but not sufficient restriction on the parameter $\al$ follows from it, namely,
$\al^2\le |\La|/3$. Seeking the sufficient condition in this case turns out to be a very complicated problem, currently unsolved. Therefore, the problem of the existence of a global spiral embedding for the RN-AdS black hole remains open.

As a result, we conclude that minimal global spiral embeddings exist for the S, S-dS, RN, and,
probably, RN-AdS black holes.

\subsection{Exponential embeddings}\label{expo}
Using (\ref{1}), we calculate the metric components
\disn{21}{
g_{00}=-2\al f+\dz k^2,\qquad
g_{01}=f'-2f w'+\dz k h',\qquad
g_{11}=\frac{2}{\al}w'\ls f'- f w'\rs+\dz h'^2-1,
\nom}
whence using (\ref{2}) and (\ref{3}), we conclude that
\disn{22}{
f=-\frac{F-\dz k^2}{2\al},\qquad
w'=\frac{F'-2\dz\al k h'}{2(F-\dz k^2)}.
\nom}
As in the case of spiral embeddings, the smoothness in terms of the functions $h(r)$ and $w(r)$ cannot be
verified (see the remark after (\ref{19})). Therefore, it is convenient to write the equation arising from formula (\ref{1})
for the component $g_{11}$ in terms of the function
\disn{22.1}{
u(r)=w(r)-\frac{\al}{k} h(r),
\nom}
which must be smooth for smooth embeddings. We write the solution of this quadratic equation for $u'(r)$immediately:
 \disn{23}{
u'=\frac{k^2 F'\pm\sqrt{\dz k^2\ls F F'^2+4\al^2(F -\dz k^2)(F-1)\rs}}{2 k^2(F -\dz k^2)}.
\nom}
t is easy to see that if $F(r)-\dz k^2$ vanishes at a certain point  $r_*$, then the expression in the right-hand
side of (\ref{23}) is smooth for a certain choice of the sign of the root: this sign must be opposite to the sign
of $F'(r_*)$. Because the sign of the root must be chosen the same for all $r$, this means that the embedding
must be smooth everywhere only for a choice of the parameter $k$ such that $F'(r)$ has the same sign at all
points where $F(r)=\dz k^2$. This condition is used below.

It is also necessary for smoothness that the radicand in (\ref{23}) (we let $P(r)$ denote it) be nonnegative
everywhere. Because the minimal global embeddings for the S and RN black holes were respectively
constructed in \cite{statja27} and \cite{statja30}, we analyze the existence of such embeddings in the remaining four cases. We
write the asymptotic form of the radicand in (\ref{23}) for them. As $r\to0$, we have
 \disn{24}{
\text{S-dS, S-AdS:}\quad P(r)\str{r\to0}-8\dz\frac{k^2m^3}{r^5},\qquad
\text{RN-dS, RN-AdS:}\quad P(r)\str{r\to0} 4\dz\frac{k^2q^6}{r^8},
\nom}
and as  $r\to\infty$  in all four cases, we obtain
 \disn{25}{
P(r)\str{r\to\infty}\dfrac{4}{27}\dz k^2 \La^2\ls 3\al^2-\La\rs r^4+\dfrac{4}{9}k^2\La\ls 3\al^2 k^2-\dz\ls 3\al^2-\La\rs\rs r^2+\ldots
\nom}
(we recall that  $\La>0$ in the dS cases and $\La<0$ in the AdS cases). Comparing formulas (\ref{24}) and (\ref{25}) shows
that it is impossible to provide the nonnegativity of the radicand for the S-AdS black hole for any choice of
$\dz$, and this means that a global embedding cannot be constructed. We must take $\dz=-1$ and $\al\le\sqrt{\La/3}$
in the S-dS case, $\dz=+1$ in the RN-dS and RN-AdS cases, and $\al\ge\sqrt{\La/3}$ in the RN-dS case. In the S-dS
case with $\dz=-1$ and the RN-AdS case with $\dz=+1$, we can see that $F(r)-\dz k^2$ must vanish twice at
points where $F'(r)$ has different signs (see Fig. 1). Therefore, global embeddings cannot be constructed in
these cases (see the remark after formula (\ref{23})).
As a result, a possibility of constructing a global embedding remains only in the RN-dS case with
$\dz=+1$ and $\al\ge\sqrt{\La/3}$. If we additionally set $k=1$, then the radicand becomes
 \disn{26}{
P(r)=F F'^2+4\al^2(F-1)^2.
\nom}
For the function $F(r)$ corresponding to the RN-dS case (see (\ref{3})), we can choose a rather large value of $\al$
such that expression  (\ref{26}) is positive for any $r$ (although it is nontrivial to precisely estimate the required
value of $\al$). Consequently, the minimal global exponential embedding for the RN-dS black hole can be
represented in the form
 \disn{27}{
y^+=\frac{1}{\al}\exp(\alpha \widetilde t+u(r)),\qquad
y^-=\frac{1-F(r)}{2\al}\exp(-(\alpha \widetilde t+u(r))),\qquad
y^2= \widetilde t
\nom}
(the remaining components $y^{3,4,5}$ are the same as in (\ref{4})), where
 \disn{28}{
u(r)=\int\! dr\,\frac{F'+\sqrt{F F'^2+4\al^2(F-1)^2}}{2 (F -1)},
\nom}
$\widetilde t=t+h(r)$  and h(r) has the form of an integral that can be easily written using formulas (\ref{22}), (\ref{22.1}) and (\ref{28}).

As a result, we conclude that minimal global exponential embeddings exist for the S, RN, and RN-dS
black holes.

\subsection{Cubic embeddings}\label{kub}
Using (\ref{1}), we calculate the metric components
\disn{29}{
g_{00}=2\dz\al f,\qquad
g_{01}=\dz (2\al f h'+u'),\qquad
g_{11}=\dz(2\al f h'^2-f'^2+2h' u')-1,
\nom}
whence using (\ref{2}) and (\ref{3}), we conclude that
\disn{30}{
f=\dz\frac{F}{2\al},\qquad
h'=-\dz\frac{u'}{F}.
\nom}
As can be seen from (\ref{9}), for a smooth embedding, the function $u(r)$ must be smooth. It is therefore
convenient to write the equation arising from (\ref{1}) for the component $g_{11}$ in terms of precisely this function.
Such an equation has the form
 \disn{31}{
u'^2=1-F-\dz\frac{FF'^2}{4\al^2}.
\nom}
For the smoothness of $u(r)$ and consequently the entire embedding, the right-hand side of this equation
must be nonnegative for any $r$. The minimal global embeddings for the S, S-dS, and RN black holes were
respectively constructed in \cite{statja27}, \cite{statja32} and \cite{statja30}.
 It only remains for us to study the possibility of constructing
such embeddings in the remaining three cases. We write the asymptotic forms of the right-hand side of (\ref{31})
for them. As $r\to 0$, we have
 \disn{32}{
\text{S-AdS:}\quad u'^2\str{r\to 0} 2\dz\frac{m^3}{\al^2 r^5},\qquad
\text{RN-dS, RN-AdS:}\quad u'^2\str{r\to 0}-\dz\frac{q^6}{\al^2 r^8},
\nom}
and as $r\to\infty$  in all three cases, we have
 \disn{33}{
u'^2\str{r\to \infty}\dz\frac{\La^3 r^4}{27\al^2}.
\nom}
Comparing these formulas, we can easily see that for the S-AdS and RN-dS black holes, it is impossible
to provide the nonnegativity of the right-hand side of (\ref{31}) for any choice of $\dz$, which means that global
embeddings cannot be constructed. In the remaining RN-AdS case, such a nonnegativity in the limits $r\to0$
and $r\to\infty$ exists for $\dz=-1$. But it is still unclear whether the choice of the parameter $\al$ can provide it
for all $r$; this problem requires an additional study.

As a result, we conclude that minimal global cubic embeddings exist for the S, S-dS, RN, and,
probably, RN-AdS black holes.

\section{Conclusions}\label{zakl}
The results of our analysis concerning the existence of minimal global embeddings for nonrotating
black holes are given in Table \ref{tab2} (for embedding types not given in Table \ref{tab2}, there are no global embeddings;
see the beginning of Sec. \ref{analiz}).
\begin{table}[htbp]
\begin{center}
\begin{tabular}{|c|c|c|c|c|c|c|}
\hline
                  &      \zagt{S}     &    \zagt{S-dS}    &\zagt{S-AdS}&     \zagt{RN}     &\zagt{RN-dS}&\zagt{RN-AdS}\\
\hline
hyperbolic   & 1+5,\cite{frons}  &      ---          &    ---     &        ---        &    ---     &     ---     \\
\hline
spiral        &1+5,\cite{statja27}&1+5,\cite{statja32}&    ---     &2+4,\cite{statja30}&    ---     &   2+4,\;\;? \\
\hline
exponential  &1+5,\cite{davidson}&      ---          &    ---     &2+4,\cite{statja30}&   2+4      &     ---     \\
\hline
cubic       &1+5,\cite{statja27}&1+5,\cite{statja32}&    ---     &2+4,\cite{statja30}&    ---     &   2+4,\;\;? \\
\hline
\end{tabular}
\end{center}
\caption{Global minimal embeddings of nonrotating black holes}
\label{tab2}
\end{table}
By globality, we mean that the embedding is smooth for all radius values, and
by minimality, we mean the minimal possible dimension of the ambient space, which is six in this case.
Table 2 contains the signature of the ambient space for each known embedding (the number of timelike
directions plus the number of spacelike directions) and a reference to the paper where this embedding was
first constructed (if no reference is given, then it was constructed here). Dashes denote the cases where the
corresponding embedding does not exist, and question marks denote the cases where the question of the
existence of the embedding remains open.

To obtain the answer to the question whether the embeddings exist in the remaining cases, it is
necessary to study whether some high-degree polynomials (in the right-hand sides of Eqs. (\ref{20}) and (\ref{31}))
vanish in certain intervals. For this, the method of complete root classification (CRC), whose description can be
found, e.g., in \cite{Liang20091487}, can be useful. In particular, it allows obtaining restrictions imposed on the coefficients of
the parametric polynomial for which it has no negative roots. This method can be successfully algorithmized,
but it requires significant computational facilities.

{\bf Acknowledgements.}
 The authors are grateful to the organizers of the conference ''In Search of Fundamental Symmetries''{} dedicated to the 90th birthday of Yu.~V.~Novozhilov and also expresses their gratitude
to A.~Erokhin for the given references. The work was supported by Saint Petersburg State University grant N~11.38.223.2015.


\end{document}